\documentstyle[12pt,twoside]{article}
\begin{document}
\begin{center}

{\Large\bf REGULAR BOUNCING\\[5PT]
COSMOLOGICAL SOLUTIONS IN\\[5PT]
EFFECTIVE ACTIONS IN FOUR DIMENSIONS\\[5pt]}
\medskip
 
{\bf C. P. Constantinidis\footnote{e-mail: clisthen@cce.ufes.br},
J. C. Fabris\footnote{e-mail: fabris@cce.ufes.br}, R. G. Furtado 
\footnote{e-mail: furtado@cce.ufes.br} and M. Picco\footnote{e-mail:
picco@lpthe.jussieu.fr} 
\footnote{Permanent address: LPTHE, Universit\'e de Paris VI et Paris VII, 
4 Place Jussieu, 75252 Paris Cedex 05,
France.}} \medskip

Departamento de F\'{\i}sica, Universidade Federal do Esp\'{\i}rito Santo, 
29060-900, Vit\'oria, Esp\'{\i}rito Santo, Brazil
\medskip

\end{center}
 
\begin{abstract}
We study cosmological scenarios resulting from effective actions in four
dimensions which are, under some assumptions, connected with
multidimensional, supergravity and string theories. These effective
actions
are labeled by the parameters $\omega$, the dilaton coupling constant, and
$n$ which establishes the coupling between the dilaton and a scalar
field originated from the gauge field existing in the original theories.
There is a large class of bouncing as well as Friedmann-like solutions. We
investigate under which conditions bouncing regular solutions can be
obtained. In the case of the string effective action, regularity is
obtained through the inclusion of contributions from the Ramond-Ramond
sector of superstring.
\vspace{0.7cm}

PACS number(s): 04.20.Cv., 04.20.Me
\end{abstract}
 
Higher dimensional space-time, supersymmetry and strings are some of the
most outstanding concepts employed currently for the construction of a
theory unifying all physical interactions.  The emergence of the idea of a
string replacing point particles as the fundamental element in nature has
permitted one to obtain a multiplet including the graviton together with
massless spin $1$ particles \cite{green}.  Supersymmetry is naturally
included in it, preserving the unitarity and avoiding tachyons, and
completing the fermionic sector of the multiplet. The critical dimension
of
the bosonic string is $26$, while its
supersymmetric version, the superstring, has a critical dimension equal to
$10$. In this way, supergravity and Kaluza-Klein theories are, in some
sense, inserted in the framework of this unification program. The true
nature of superstring theories is revealed on very high energy levels,
near
the Planck scale, $E_p \sim 10^{19} GeV$. Hence, the early Universe is the
most viable laboratory to test this theory, unless some phenomenology can
be obtained at a much smaller energy scale, which is not the case until
now.  
\par The structure of string theory is very complex and in trying to
study its cosmological consequences, it comes out more feasible to work
with the effective theory in the low energy level \cite{green,kiritsis}.
Even at this level the theory presents a non-trivial field composition:
essentially, there is the gravity sector, the dilaton and a
three-form field. In the gravity sector, at one loop approximation, the
Gauss-Bonnet
term comes to scene. The dilaton field couples non-minimally to gravity.
All these terms are written in ten dimensions and the reduction to four
dimensions leads to the appearance of moduli fields via the
compactification process.  
\par
These moduli fields appear also in
multidimensional theories, not necessarily related to strings. In
principle, a pure multidimensional theory must contain just geometry, the
reduction to four dimensions producing gauge and scalar fields. But, if we
consider a supergravity theory in higher dimensions (the most famous ones
being the eleven and ten dimensional supergravities), gravity, in its
higher dimensional formulation, is generally coupled to gauge or scalar
fields, composing the bosonic sector. The reduction to four dimensions
leads to a non-trivial coupling between gravity and scalar fields and
among
scalar fields themselves, some of them coming from the compactification of
the internal dimensions.  
\par 
The cosmological consequences of these effective models in four
dimensions has been extensively studied.  Some examples are the
string \cite{antoniadis,barrow,veneziano,gasperini}, multidimensional
\cite{freund} and supergravities \cite{spindel,fabris1} effective
models. In some cases, multidimensional effective models with gauge
fields have been considered, and in particular with a higher
dimensional conformal gauge field theory \cite{fabris2}.  The low
energy superstring effective model is the base of the so-called
pre-big-bang \cite{gasperini} scenario, where there is a phase, prior
to the radiative era, during which the Universe could be either in an
expanding inflationary regime or in a contraction phase, without
reaching the singularity.  
\par
The structure of the coupling between gravity and scalar
fields in these effective theories in 4 dimensions can be, under
certain assumptions to be specified latter, represented by the
following expression:
\begin{equation}
\label{l1}
{\it L} = \sqrt{-g}\biggr[\biggr(\phi R -
\omega\frac{\phi_{;\rho}\phi^{;\rho}}{\phi} - 
\phi^n\Psi_{;\rho}\Psi^{;\rho}\biggl)
- \chi_{;\rho}\chi^{;\rho}\biggl] + {\it L_m}.
\end{equation}
The term ${\it L_m}$ represents the ordinary matter. Here, we will be
interested in the radiative fluid only, since 
it seems more realistic when we have in mind the primordial Universe. As
examples of this Lagrangian, we note that: 
\begin{enumerate}
\item It corresponds to pure multidimensional theories with $\Psi = \chi
=$
constant and $\omega = (1 - d)/d$, where $d$ is the 
number of compactified dimensions (in this case, we consider the
compactification on the torus); 
\item If we consider a two form-gauge field in higher dimensions
we obtain $\Psi \neq$ constant, and $n = - \frac{2}{d} + 1$;
\item A conformal gauge field, represented by a $(d + 4)/2$-form, leads to
$n = - \frac{2}{d}$; 
\item In string theory $\omega = -1$; moreover, the fact that in general
the three-form field $H_{\mu\nu\lambda}$ has, in 
four dimensions,
just one degree of freedom permits one to write $H^{\mu\nu\lambda} =
\phi^{-1}\epsilon^{\mu\nu\lambda\sigma}\Psi_{;\sigma}$
which corresponds to $n = - 1$ in (\ref{l1}) \cite{maeda}.
\end{enumerate}
\par 
The $\chi$ term is the scalar component coming from another three-form
field which is present in the Ramond-Ramond
sector of type IIB superstring. Note that this
field does not couple directly with the dilaton.
Since we are interested in cosmological
applications, we have set all gauge fields equal to zero in the effective
model in four dimensions. Moreover, in all these cases, there are some
simplifications in the process of obtaining the effective action. For
string theory, for example, the moduli fields and the Gauss-Bonnet term
are absent; in what concerns multidimensional models, the moduli fields
are
taken into account but the curvature of the internal space is zero.  
\par
The purpose of this paper is to consider the Lagrangian (\ref{l1}) with
$\omega$ and $n$ arbitrary.  In this way, we can map many possible
cosmological scenarios resulting from the existence of supersymmetry and
extra dimensions, with the assumptions specified above.  Our interest is
to
verify the generality of bouncing and, in some cases, inflationary
solutions.  Of course, a bouncing solution has necessarily an inflationary
phase,
since a minimum for the scale factor implies $\dot a = 0$ and $\ddot a >
0$
near it. The main question we try to answer is the conditions under which
a
bouncing or inflationary Universe may be found in the realm of the
effective
theories described by (\ref{l1}). We remark, however, that the existence
of
a bouncing solution is not a sufficient condition for having
singularity-free
models, as will be shown later.  We will analyze also the existence of
complete regular solutions.  
\par In fact, if we restrict ourselves first
to the string cosmology program, it is has already been shown
\cite{maeda,wands}
that the existence of an axion field in the tree level string effective
action leads to the existence of a minimum for the scale factor, that is a
bouncing solution. However, a singularity is still present in the
beginning
of the evolution of the Universe, even if the scale factor is non zero. In
this sense, the introduction of loop approximations is an essential
ingredient of string cosmology since it permits to avoid the
singularity. However, these analysis were made considering $\omega = - 1$
and $n = -1$, i.e. string cosmology with a particular coupling between the
dilaton and axion fields. We will extend them here, showing that for
$\omega = - 1$ a singularity seems to be unavoidable. To avoid this
singularity, we must have $n < 1$ and $- \frac{3}{2} > \omega > -
\frac{4}{3}$.  In string cosmology, this implies to consider
moduli fields, whose analysis is more complex since a third scalar field,
with non trivial coupling with the others two, appears in
(\ref{l1}). But we will show that the introduction of terms coming from
the Ramond-Ramond
sector of superstring can render the bouncing solutions regular even at
the tree level.  
\par 
We
will study essentially the predictions of models described by (\ref{l1})
concerning the evolution of the scale factor in the primordial Universe.
We will verify that bouncing solutions are most favored by the conditions
$\omega < 0$ and $n < 1$, even if in some situations it can occur for
other
values of these parameters.  Complete regular solutions can be obtained
when $\omega = 0$ and $n < 1$ (which, on the other hand, seems to be
unstable),
or for $- \frac{3}{2} > \omega > - \frac{4}{3}$.  We will perform our
study in the vacuum case, where only
those fundamental fields are present, and when they are coupled to a
radiative fluid. The radiative fluid will be minimally coupled to the
geometry in Jordan's frame, where dilaton is non-minimally coupled to
gravity.  
\par The field equations coming from Lagrangian (\ref{l1}) are
\begin{eqnarray}
\label{fe1}
R_{\mu\nu} - \frac{1}{2}g_{\mu\nu}R &=& \frac{8\pi}{\phi}T_{\mu\nu} +
\frac{\omega}{\phi^2}\biggr(\phi_{;\mu}\phi_{;\nu}
- \frac{1}{2}g_{\mu\nu}\phi_{;\rho}\phi^{;\rho}\biggl) +
\frac{1}{\phi}\biggr(\phi_{;\mu;\nu} - g_{\mu\nu}\Box\phi\biggl) +
\nonumber\\
&+& \phi^{n-1}\biggr(\Psi_{;\mu}\Psi_{;\nu} -
\frac{1}{2}g_{\mu\nu}\Psi_{;\rho}\Psi^{;\rho}\biggl) 
+ \frac{1}{\phi}\biggr(\chi_{;\mu}\chi_{;\nu} -
\frac{1}{2}g_{\mu\nu}\chi_{;\rho}\chi^{;\rho}\biggl) \quad ;\\
\nonumber\\
\label{fe2}
\Box\phi &+& \frac{1 - n}{3 + 2\omega}\phi^n\Psi_{;\rho}\Psi^{;\rho} +
\frac{\chi_{;\rho}\chi^{;\rho}}{3 + 2\omega}= \frac{8\pi}{3 + 2\omega}T
\quad ; \\ 
\nonumber \\
\label{fe3}
\Box\Psi + n\frac{\phi_{;\rho}}{\phi}\Psi^{;\rho} &=& 0 \quad ; \\
\Box\chi &=& 0 \quad \quad ; \\
{T^{\mu\nu}}_{;\mu} &=& 0
\end{eqnarray}
We insert in these equations the FLRW metric,
\begin{equation}
ds^2 = dt^2 - a(t)^2\biggr(\frac{dr^2}{1 - \epsilon r^2} + r^2(d\theta^2 +
\sin^2\theta d\phi^2)\biggl) \quad . 
\end{equation}
A flat, open and closed spatial section implies $\epsilon = 0, -1, 1$,
respectively. 
We consider also a barotropic perfect fluid,
\begin{equation}
T^{\mu\nu} = (\rho + p)u^\mu u^\nu - p g^{\mu\nu} \quad , \quad p =
\alpha\rho \quad - 1 \leq \alpha \leq 1 \quad . 
\end{equation}
\par 
As it happens in general when we are treating gravity in the
presence of scalar fields, it is more convenient to reparametrize the
time coordinate such that
\begin{equation}
\label{deftime}
dt = a^3d\theta \ .
\end{equation} 
In this new parametrization and with primes denoting derivatives with 
respect to $\theta$, the equations of motion read
\begin{eqnarray}
\label{em1}
3\biggr(\frac{a'}{a}\biggl)^2 + 3\epsilon a^4 &=& \frac{8\pi}{\phi}\rho
a^6 +
\frac{\omega}{2}\biggr(\frac{\phi'}{\phi}\biggl)^2 -
3\frac{a'}{a}\frac{\phi'}{\phi} + 
\frac{\phi^{n-1}}{2}\Psi'^2 + \frac{1}{2}\frac{\chi'^2}{\phi} \quad , \\
\label{em2}
\phi'' + \frac{1 - n}{3 + 2\omega}\phi^n\Psi'^2 + \frac{\chi'^2}{3 +
2\omega} &=& 
\frac{8\pi}{\phi}(\rho - 3p)a^6 , \quad , \\
\label{em3}
\Psi'' + n\frac{\phi'}{\phi}\Psi' &=& 0 \quad , \\
\label{em4}
\chi'' &=& 0 \quad , \\
\label{em5}
\rho' + 3\frac{a'}{a}(\rho + p) &=& 0 \quad .
\end{eqnarray}
The equations (\ref{em3},\ref{em4},\ref{em5}) are easily integrated,
leading to 
\begin{equation}
\label{fi1}
\Psi' = A\phi^{-n} \quad , \quad \chi = B\theta +C \quad , \quad \rho =
\rho_0a^{-3(1 + \alpha)} \quad , 
\end{equation}
where $A, B, C$ and $\rho_0$ are integration constants.
The integration of the other two equations depends on whether $\chi =$
constant or not and on the 
presence of ordinary matter. We will consider separately these different
cases. 

\vspace{0.3cm}

\noindent
{\bf I)} Vacuum case, $\rho=0$.
\noindent

\vspace{0.3cm}

\vspace{0.3cm}

\noindent
{\bf Ia)} $\chi =$ constant.
\noindent

\vspace{0.3cm}
 
In this case, equation (\ref{em2}) admits a first integral,
\begin{equation}
\label{fi2}
\frac{\phi'^2}{2} + \frac{A^2}{3 + 2\omega}\phi^{1-n} = \frac{D}{2} \quad
,
\end{equation}
where $D$ is an integration constant. From (\ref{deftime},\ref{fi2}) we
obtain a
transcendental 
relation for $\phi$ as function of $\theta$, through hypergeometric
functions 
\begin{equation}
\theta = - \frac{2}{(n-1)\sqrt{D}}\phi\;{_2F_1}(\frac{1}{1 -
n},\frac{1}{2},\frac{2 - n}{1 - n},(r\phi)^{1-n}) \quad , 
\end{equation}
where $r = \biggr(\frac{(3+2\omega)D}{2A^2}\biggl)^2$.
In order to integrate (\ref{em1}), we redefine the scale factor $a =
\phi^{-1/2}b$, 
obtaining the equation
\begin{equation}
\label{eq18}
3\biggr(\frac{b'}{b}\biggl)^2 + 3\epsilon \frac{b^4}{\phi^2} =
\frac{3+2\omega}{4}\biggr(\frac{\phi'}{\phi}\biggl)^2 +
\frac{A^2}{2}\phi^{-1-n} 
\quad .
\end{equation}
We will first study the cases where $n \neq 1$; the case $n = 1$ will be
studied separately. 
Using (\ref{fi2}), equation (\ref{eq18}) becomes
\begin{equation}
\frac{b'}{b} = \biggr(\frac{3+2\omega}{12}D - \epsilon
b^4\biggl)^{1/2}\frac{1}{\phi} \quad . 
\end{equation}
Eliminating again $d\theta$ through (\ref{fi2}), one obtains the integral
relation 
\begin{equation}
\label{ir}
\int \frac{dy}{y\sqrt{1 - \epsilon y^2}} =
\frac{2}{1-n}\sqrt{\frac{3+2\omega}{3}}\int \frac{dx}{x\sqrt{1 - x^2}}
\quad , 
\end{equation}
where $x = \sqrt{\frac{2A^2}{(3+2\omega)D}}\phi^{(1-n)/2}$ and $y =
\sqrt{\frac{12}{(3 + 2\omega)D}}b^2$.  In order to impose that $x^2 <
1$, we define $x= \sin\xi$.  Integrating (\ref{ir}) and using the
definitions given above, we obtain the following solution for $a$:
\begin{equation}
\label{sf1}
a = a_0(\sin\xi)^{1/(n-1)}\biggr[\frac{\tan^p\frac{\xi}{2}}{1 +
\epsilon k^2\tan^{2p}\frac{\xi}{2}}\biggl]^{1/2} \quad ,
\end{equation}
where $k$ is another integration constant and $p = \frac{2}{1 -
n}\sqrt{1 + \frac{2}{3}\omega}$.  The cosmic time is connected with
the variable $\xi$ through the relation
\begin{equation}
t = \frac{L}{1 - n}\int a^3\sin^\frac{1+n}{1-n}\xi d\xi \quad ,
\end{equation}
where $L = 2\biggr(\frac{2A^2}{(3+2\omega)D}\biggl)^{1/(1-n)}$. The
integrand of this relation being always positive, $t$ is a monotonic
function of $\xi$.
\par 
For $\epsilon = 0$, the asymptotic behavior for $\xi \rightarrow
0$ is $a \propto \xi^{\frac{1}{1-n}(-1 + \sqrt{1 +
\frac{2}{3}\omega})}$, while for $\xi = \pi - z$, $z \rightarrow 0$,
the asymptotic behavior reads $a \propto z^{\frac{-1}{1-n}(1 + \sqrt{1
+ \frac{2}{3}\omega})}$. From these expressions, we can classify the
possible scenario: for $n < 1$, $\omega < 0$, (\ref{sf1}) represents a
bouncing solution; for $n > 1$ and $\omega < 0$, we find a big bang
followed by
a big crunch, in spite of the fact that the spatial section
is flat; all other cases represent an expanding universe.
\par
Following the same asymptotic analysis we find bouncing solutions
for $\epsilon \neq 0$ only when $\omega < 0$ and $n < 1$. For the other
cases,
the corresponding Friedmann-like scenarios are recovered.
\par 
A special case is $\omega = 0$. Then, the solution (\ref{sf1})
takes the form
\begin{equation}
\label{oscila}
a = a_0\frac{1}{\sqrt{\cos^{4/(1-n)}\frac{\xi}{2} + \epsilon
k^2\sin^{4/(1-n)}\frac{\xi}{2}}} \quad .
\end{equation}
In this case, if $n < 1$, we are at the boundary of two different
behaviors and this special solution reflects this fact.  For
$\epsilon = 0$, $n < 1$, (\ref{oscila}) is a sequence of bouncing
Universes. If $\epsilon = - 1$ we find again a bouncing Universe. When
$\epsilon = 1$ the Universe oscillates, the scale factor never
reaching zero. This last case is an example of a complete, eternal,
regular Universe. In particular, when $k = 1$, we find a static
Universe even if the scalar fields evolve with time.  If $n > 1$, on
the other hand, (\ref{oscila}) represents, for $\epsilon = 0, 1$, an
oscillating Universe, the minimum value of the scale factor being
zero. For $\epsilon = - 1$ we find again a bouncing Universe.
\par
There exists also the limit case $n = 1$ for which the scalar field
behaves as $\phi = E\theta$, $E$ being a constant. The equation for
the scale factor can be integrated in a similar way as before, leading
to the following expressions:
\begin{eqnarray}
\epsilon &=& 1, -1 \quad , \quad a = a_
0\theta^{-1/2}\sqrt{\frac{\theta^{\pm q}}{1 + \epsilon\theta^{\pm
2q}}} \quad ,\\ \epsilon &=& 0 \quad , \quad a = a_0 t^\frac{q \pm
1}{3q \pm 1} \quad , \\
\end{eqnarray}
where $q = \sqrt{1 + \frac{2}{3}\omega + A^2}$. These solutions
represent an expanding Universe with an initial singularity, for any
value of the curvature constant $\epsilon$.  There are inflationary
expanding solutions when $- \frac{3}{2} - A^2 < \omega < - \frac{4}{3}
- A^2$.

\vspace{0.3cm}

\noindent
{\bf Ib)} $\chi \neq$ constant.
\noindent

\vspace{0.3cm}

This case seems to admit an exact solution only for $n = - 1$.
The first integral for $\Psi$ and $\chi$ are
\begin{equation}
\Psi' = A\phi \quad , \quad \chi = E\theta \quad .
\end{equation}
Using these first integrals, the solution for the scalar field 
$\phi$ reads
\begin{equation}
\phi = C\sin\kappa\theta - \frac{E^2}{2A^2} \quad , \kappa =
\sqrt{\frac{2A^2}{3 + 2\omega}} \quad ,
\end{equation}
while the equation for the scale factor can be reduced to
\begin{equation}
\int \frac{dy}{y\sqrt{1 - y^2}} = \frac{2D}{\sqrt{3}}\int
\frac{d\theta}{\phi} \quad ,
\end{equation}
where $a = \phi^{-1/2}b$, $y^2 = \sqrt{\frac{3\epsilon}{D^2}}b^4$.
The final solutions are:
\begin{eqnarray}
\label{cs1}
\epsilon &=& 1 \quad , \quad a = a_0\frac{1}{\sqrt{\sin\kappa\theta -
s}}\cosh^{-1/2}f(\theta) \quad ,\\
\label{cs2}
\epsilon &=& 0 \quad , \quad a = a_0\frac{1}{\sqrt{\sin\kappa\theta -
s}}\exp(f(\theta)) \quad , \\
\label{cs3}
\epsilon &=& - 1 \quad , \quad a = a_0\frac{1}{\sqrt{\sin\kappa\theta
- s}}\sinh^{-1/2}f(\theta) \quad ,
\end{eqnarray}
where
\begin{equation}
f(\theta) = \frac{4D}{\sqrt{3}C\kappa}\frac{1}{\sqrt{\mid s^2 -
1\mid}}\arctan\biggr[\frac{1 - s\tan\frac{\kappa\theta}{2}}{\sqrt{\mid
s^2 -1\mid}}\biggl] \quad , \quad s = \frac{E^2}{2A^2C} \quad .
\end{equation}
These solutions are valid for $s < 1$, which is the most interesting
case.  The solutions (\ref{cs1},\ref{cs2},\ref{cs3}) represent a
bouncing Universe for any value of $\omega$.  Moreover, there is a
local maximum for the scale factor when $\epsilon = 1$ and $\theta =
\frac{2}{\kappa}\arctan(1/s)$.

\vspace{0.3cm}

\noindent
{\bf II)} Radiative fluid case and $\chi = $ constant.
\noindent

\vspace{0.3cm}

The integration of the equations follows the same procedure as before. We
note that for the radiative case 
${T^\rho}_\rho = T = 0$. Hence the equations for the scalar fields, and
their corresponding first integrals, keep their 
form. However, for the scale factor, and after redefining $a =
\phi^{-1/2}b$, we obtain 
\begin{equation}
\frac{b'}{b} = \biggr(\frac{3+2\omega}{12}D + m b^2- 3\epsilon
b^4\biggl)^{1/2}\frac{1}{\phi} , \quad 
\end{equation}
where we have written $m = 8\pi\rho_0$.
The first integral of (\ref{em2}) permits again to employ the
reparametrization $\phi^{1 - n} = \sin\xi$. We can then integrate the
equations for $b$, reconstructing the solutions 
for $a$. We will present the solution for the case $\epsilon = 0$. It
reads:
\begin{eqnarray}
a = a_0(\sin\xi)^{1/(n-1)}\biggr\{\frac{\tan^p(\xi/2)}{1 -
\tan^{2p}(\xi/2)}\biggl\} \quad , 
\end{eqnarray}
where $p$ is defined as before.
This solution exhibits bouncing solutions for $\omega < 0$ and $n < 1$.

\vspace{0.3cm}

The different scenarios exposed here have many interesting features that
deserve a
more detailed analysis, mainly through the construction of realistic
models and testing them against observations.  We studied the
behavior of cosmological models inspired in string, supergravities and
Kaluza-Klein theories. These different effective models can be recast in a
unified form, labelled essentially by the two parameters $n$ and
$\omega$. In what concerns the effective model coming from string there
are
two important simplifications: the Gauss-Bonnet term is not included; the
compactification from ten to four dimensions is performed on the torus,
with a constant internal scale factor. Concerning effective models coming
from multidimensional theories, one of the most important restriction is
the absence of curvature in the internal space.  
\par
The solutions obtained reveal bouncing, in general, for $n < 1$ and
$\omega < 0$. The
presence of the term $\chi$, originating from the Ramond-Ramond sector,
leads in general to the presence of a bounce. However, we must remark that
the
fact that the scale factor goes to an infinite value in the initial state
of the Universe does not mean an absence of singularity.  To verify this,
we investigate the behavior of a curvature scalar, like $R$, the Ricci
scalar. It writes as
\begin{equation}
R = - 6\biggr(\frac{\ddot a}{a} + (\frac{\dot a}{a})^2\biggl) \quad ,
\end{equation} 
and when $a \propto t^m$  (what can be an
approximation for the solutions founded before for small intervals of
time),
$R \propto t^{-2}$. Hence, $R \rightarrow \infty$
if the initial states occurs at $t = 0$. 
It is regular, if $t \rightarrow - \infty$, meaning a singularity-free
Universe.
For example, in the vacuum and radiative cases, with $\chi =$
constant, and $\epsilon = 0$, the Universe is completely regular for
$n < 1$ and $- \frac{3}{2} < \omega < - \frac{4}{3}$. When $\chi \neq$
constant, with $n = - 1$, the solutions are always regular. Hence, in
order to have a
bouncing regular solution, where the Universe begins and ends in a
Minkowskian state, in the context of string ($\omega = - 1$), the
field $\chi$ must be considered. Otherwise, there is a singular
initial state and the corrections coming from the one loop
approximation must be taken into account.
\par
Regular bouncing solutions are not so common in the literature.
In principle, one may think that the existence of a bounce by
itself can lead to singularity-free cosmological models.
But, this is not true, as it has beem shown in \cite{fabris2,maeda} for
example. However, such kind of cosmological models, if they are regular
everywhere,
represent an
interesting alternative, or amending, to the standard one since: 1) they
are free of an initial
singularity, which is one of the most important problems in the standard
cosmological
model; 2) they contain naturally an inflationary phase, that ends at a
given moment
after the begining of the expansion phase; 3) they can join the radiative
phase of
the standard model, keeping all of its advantages from the observational
point
of view. In principle, a bouncing Universe must constitute a
primordial scenario, and perhaps we must consider the possibility of
a decaying of the primordial scalar fields, originating ordinary matter,
from where a smooth transition to the ordinary radiative phase can
be obtained. But, this leads to consider a more complicate scenario
(from technical and conceptual point of view) of baryogenesis.
\par
The introduction of a radiative fluid may permit one to obtain some
observational constraints on these models. In this case, it is
possible to associate a temperature that varies as the inverse of the
scale factor, as usual. Hence, the bouncing Universe has a cold origin
and the temperature mounts to a maximum value, decreasing from that
moment on. This can leave traces in the primordial nucleosynthesis
with specific predictions for the relative abundances of light
elements and on the anisotropy of the cosmic microwave
background. The determination of these observational traces are
outside the scope of the present work, but evidently it will
constraint the parameters $\omega$ and $n$.
\par
Another good criterion to select viable models is the
stability against small perturbations.  Due to the complexity of the
background solutions, this is a very difficult analysis to be
performed. However, we note that the field $\phi$ plays the role of a
variable gravitational constant and in general there is an instability
when there is an anti-gravity phase, i.e., when $\phi$ takes negative
values during a certain period \cite{fabris2,starobinski} or when it takes
a zero value in a finite proper
time. In the
solutions determined before, the gravitational coupling is always
positive and this kind of problem is absent, excepting for some values
of the parameter $n$ when $\omega = 0$, where the gravitational can become
negative or zero during the evolution of the Universe. Hence, the vacuum
case
with $\omega = 0$ must be unstable.
\par
It is important to notice that in the case of a bouncing Universe,
there is a primordial phase where not only the strong energy condition
can be violated (leading to an inflationary phase), but also in some
cases the weak energy condition\cite{visser}.  However, there are many
examples of physical systems that can violate the weak energy
condition under certain circumstances \cite{visserbook} and this can
not be viewed as a drawback of the models exhibiting a bounce, mainly
when these models concern a primordial phase.
\par
Finally, we must stress the fact that we have worked here in the
Jordan's frame. However, the solutions can also be written in the
Einstein frame, since the redefinition $a = \phi^{-1/2}b$ employed to
solve the equations is just a conformal transformation that transports
the action from one frame to another. It is important to remark that,
in Einstein frame, all solutions have a singular initial state. This
leads us to the problem of what is the physical frame
\cite{gunzig,levin1,levin2}. Our choice was to work in the Jordan
frame since the effective model appears naturally in it.
\vspace{0.5cm}
\newline
{\bf Acknowledgement:} We thank CNPq and CAPES (Brazil) for financial
support.

\end{document}